# The Electrical Property of Large Few Layer Graphene Flakes Obtained by Microwaves Assisted Exfoliation of Expanded Graphite.

Azhar A. Pirzado[a,b], Guillaume Dalmas[a], Lam Nguyen-Dinh[a,c], Ivan Komissarov[b], Francois Le Normand[b] and Izabela Janowska[a]*

[a]Institut de Chimie et Procédés pour l'Énergie, l'Environnement et la Santé (ICPEES), CNRS UMR 7515-University of Strasbourg, 25 rue Becquerel 67087 Strasbourg, France; [b]Laboratoire des sciences de l'Ingénieur, de l'Informatique et de l'Imagerie (ICube), UMR 7357, CNRS, University of Strasbourg, 23 rue du Loess, 67037 Strasbourg, France; [c]Da-Nang University of Technology, University of Da-Nang, 54 Nguyen Luong Bang, Da-Nang, Viet Nam

corresponding author : janowskai@unistra.fr

**Abstract:** Few layer graphene (FLG) was synthesized by μ-wave assisted exfoliation of expanded graphite in toluene with an overall yield from c.a. 7% to 20%. A significant difference in the absorption of μ-waves by the expanded graphite and toluene allowed a rapid heating of the medium. The number of FLG sheets varies from 3 to 12, while the lateral size of the sheets exceeds few μms. The obtained FLG exhibits very low resistance with average value of 1.6 kΩ (500 Ω minimum) which is comparable to that of high quality graphenes synthesized by CVD methods, and lower than numbers of exfoliated graphenes.

## 1. INTRODUCTION

Depending on the envisaged applications many different synthesis methods of graphene/few layer graphene (FLG) have been investigated and developed. The general aim is to produce graphene with desirable properties by possibly inexpensive and facile way. Number of application fields such as electronic, optoelectronic, etc… requires a graphene material with adequate quality, i.e. high conductivity, low sheets number and large lateral size. The latter involves individual flakes or films, where large flakes permit in an easy manner to extend the conductive films due to reduced grain boundaries number within a given film surface. The lowest sheet resistance for single suspended graphene and CVD-few layer graphene, of 28 and 10Ω/sq respectively, were reported. Once supported, the sheet resistance increased accordingly to 6.45 kΩ/sq and 1.2 kΩ/sq [1, 2]. Study of the electrical transport in individual FLG flake with different C/O ratio was performed by two-probe technique [3]. Due to the hopping mechanism of charge transport, the nonlinear I/V curve was measured for the flake with 12% (at.) of oxygen. FLG containing 4% of oxygen exhibited almost linear I/V curve and sheet resistance of 21 kΩ/sq, while 0.3% oxygen- FLG showed ohmic behavior and sheet resistance of 5 kΩ/sq. Most of reported charge transport properties however are investigated for the assembly of graphene flakes, i.e. papers, films or powders. The common values of sheet resistance for FLGs synthesized by CVD methods are 0.77 -1 kΩ/sq with few exceptions, e.g. 0.28 kΩ/sq, 0.23 kΩ/sq [4-7]. The graphene/FLG obtained by exfoliation of graphite based materials usually shows reduced conductivity compared to CVD graphenes; while harsh production and reduction conditions are required in the case of GO originated graphene. The usual conductivity of free standing papers or powders from exfoliated graphenes oscillates at $10^2$ S/m with maximum value of few $10^4$ S/m and sheet resistance of few-dozen kΩ [8-10]. The exceptionally high conductivity, i.e. $10^5$ S/m, was obtained with a rapid exfoliation in a pre-heated furnace, while the best exfoliation efficiency was defined for low lateral size starting graphite [11]. Other rapid exfoliation of GO was carried out in μ-waves furnace giving material with final conductivity of 274 S/m [12]. Concerning the μ-waves assisted exfoliations only few works have been reported up to now, and electrical measurements of the final materials have been performed scarcely [13, 14].

At present we report on the synthesis and charge transport property of graphene/FLG flakes, which are obtained by μ-waves assisted exfoliation of expanded graphite in toluene. The chemical and morphological properties characterized by TEM, Raman and XPS spectroscopy reveal high quality graphene, which is finally reflected by a very low resistance measured by the two-probe method.

## 2. MATERIALS AND METHOD

### 2.1. Materials

The expanded graphite (EG) was purchased from Carbone Lorraine. Toluene was purchased from Sigma Aldrich (98.8% pure) and was used as received.

The FLG was synthesized as follows: 400 mg of EG was shortly ultrasonicated in 200 ml of toluene in a low-power sonic bath sonicator (Elma sonic 102168022) and a tip sonicator (Bandelin Sonoplus uw 2200) at 10% power, for 5 min. each or stirred in the closed vessel for 24h after bath ultrasonication. The microwave experiment was carried out in a Mars (CEM Corp.) microwave oven in four reactors at 600 W of power and 160°C for 10 min. The suspension was cooled for 15min, and the FLG sheets were extracted from the supernatant formed above the precipitated residual EG after decantation for 90 min. The final material was dried in oven at 120°C for 20h.

*2.2. Calculations*

Mass of toluene:

$m_{toluene}$ = 0.04335 kg ( = d × V = 867 kg/l × 50 × $10^{-3}$l), (1 reactor)

Specific heat capacity of toluene:

$C_{toluene}$ = 1.717 kJ/kgK (liquid phase, no evaporation) [15]

Final and initial temperatures of reaction media ($T_f$, $T_i$):

$T_{ftoluene}$ = 310K, $T_{fEG-toluene}$ = 438K, $T_{fgraphite-toluene}$ = 335K

$T_i$= 293K

Max. μ-waves energy supplied to 1 reactor:

E (J) = P (Watt) ×time (s)/4; P = 600W, time= 600s

E = 90000J = 5.6 × $10^{23}$eV (1 eV = 1.60217657 × $10^{-19}$J)

*2.3. Characterization Tools*

The transmission electron microscopy (TEM) was performed on Topcon 002B microscope working at 200 kV accelerated voltage with a point-to-point resolution of 0.17 nm. The FLG was dispersed prior to analysis in ethanol and dropped on a perforated carbon coated copper grid.

The scanning electron microscopy (SEM) was carried out on a Jeol JSM-6700F working at 3 kV accelerated voltage, equipped with a CCD camera.

XPS analyses was performed with a MULTILAB 2000 (THERMO) spectrometer equipped with Al K α anode (hν= 1486.6 eV).

Raman spectroscopy was performed using Horiba Scientific Labram Aramis Raman Spectrometer (JobinYvon technolgy) with the following conditions: laser wavelength of 532.15 nm, D2 filter (1% power) and spectrum recorded in the regions from 1250 to1650 and from 2600 to 2800 $cm^{-1}$, with an integration time of 100s for each region. The spectra were performed on $SiO_2$/Si wafer supported flakes.

The electrical measurements were performed on Karl Suss PM8. The FLG was dispersed in ethanol and deposited on interdigitated FET-like device with gold electrodes with gaps of 2.5, 5 and 10 μm respectively. Deposited flakes were annealed prior to the measurements in Ar.

### 3. RESULTS AND DISCUSSIONS

The exfoliation of EG is initially performed by submission of EG suspension in toluene with a concentration of 2 mg/ml to μ-waves irradiation at 600W power for 10 min. followed by a cooling step. The resulting product consists of settled down EG and toluene supernatant. The latter contains exfoliated (extracted) few layer graphene flakes (FLG), which are next separated with a yield of ~ 7%. Similar yield was reached earlier for the EG exfoliation in ammonia, where two gaseous molecules were produced during ammonia thermal decomposition [13]. At present however, a stirring step (24 h, closed vessel) is performed just after a short sonication and prior to the μ-waves treatment. This step helps the intercalation of toluene and increases the yield up to 20%. The final exfoliation process is also attributed to difference in μ-waves absorption between EG and toluene. Contrary to toluene (low dielectric constant ($ε_0$)), carbon materials are known to be good absorbers of μ-waves [16]. In this regard, we apply a short temperature ramp with high μ-waves power, in order to rapidly reach high temperature in the reaction media. The high temperature of the medium can be reached rapidly only by high μ-waves absorption of EG. Fig. 1A and B depict the curves of temperature-time dependence for EG-toluene, and for comparison, the graphite plates-toluene suspension and toluene alone; all recorded during a 10 min experiment with varied ramp/power parameters.

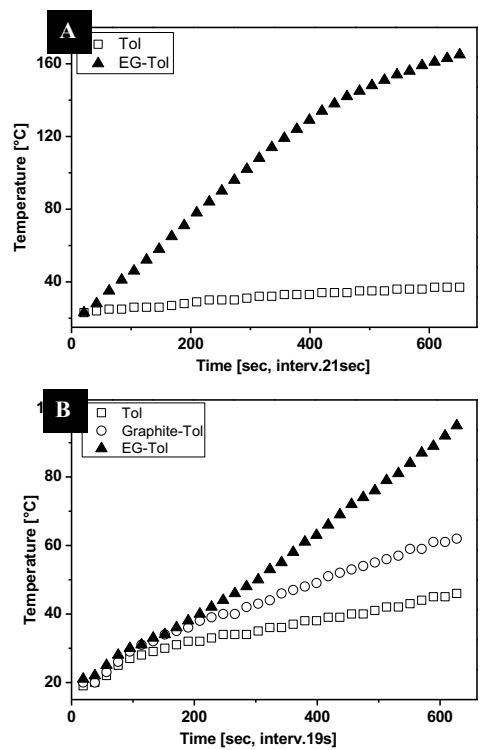

Figure 1. A and B. Temperature-time profiles obtained for toluene, EG-toluene and graphite-toluene medium under μ-waves irradiation under two different ramp/power conditions.

The curve in Fig.1A clearly shows that since toluene cannot be effectively heated by μ-waves, the significant absorption of μ-waves by EG results in very fast energy transfer and heating of toluene with a final temperature above 160°C. This rapid rise of temperature allows the EG separation (exfoliation and/or extraction), which is also favorable due to the strong electron donor-acceptor interactions of toluene and graphene through π-π stacking. Such donor-acceptor interactions can be a driving force for both, the initial intercalation of toluene between graphene sheets (especially at the cracks) and the final dispersion of the flakes in toluene. Additionally, the latter disfavors a possible collapse of exfoliated sheets during the cooling step [17]. A desorption of toluene from the graphene surface indeed requires high energy (T ≥ 450°C, at $10^{-4}$Pa) [18]. For comparison, a lower temperature of toluene and a lower heating rate is observed for graphite platelets, for which no FLG extraction is observed (Fig.1B). This is related to differences in size, surface dependent permittivity [19]. The energy absorbed by EG and transferred to Joule heat ($Q_1$), can be estimated from the difference between the energy absorbed and transferred by EG + toluene ($Q_2$) and the energy absorbed and transferred by toluene ($Q_3$) alone, according to the following equation (evaporation omitted):

$Q = m \times C \times (T_f - T_i)$, where m is the mass of toluene, C is the specific heat capacity of toluene, $T_f$ and $T_i$ are final and initial temperature, in the presence or not of EG, respectively (see *2.2*).

$Q_1$ is estimated to be 9.5 kJ (5.9…$\times 10^{22}$ eV). In the case of graphite this energy is one order magnitude lower, $6.9 \times 10^{21}$ eV, while the initial energy supplied to one reactor is estimated to be $5.6 \times 10^{23}$ eV.

A TEM analysis reveals that the number of sheets within the singular flakes does not exceed 12 with an average number of 7. Fig. 2 A-C shows representative, low and medium resolution TEM images of FLG flakes and Fig.2D the corresponding C1s XPS spectrum. A quite low calculated full width at half-maximum (FWHMs) of C1s peak, ~ 1.1eV indicates high degree of "sp$^2$" carbon. The progressive broadness of the C1s peak towards higher binding energy is linked to the proper asymmetry of sp$^2$C peak as well as the sp$^3$ C linked to oxygen groups (-OH). The tail extend corresponds to gradually increased electronegativity groups (-COH, -COOH) and shake-up line of aromatic carbon compounds at ~ 291 eV ($\pi$-$\pi$* transition). The atomic percentage of oxygen calculated from the C1s and O1s peaks areas is 5%.

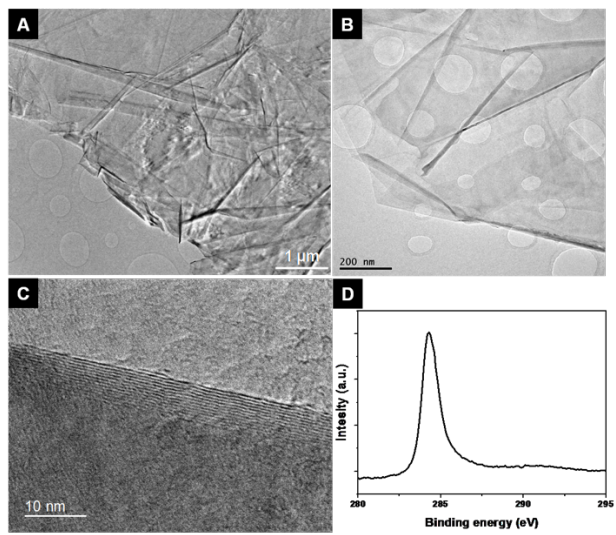

Figure 2. A-C) TEM micrographs of FLG, D) XPS C1s peak of FLG.

The Raman spectra performed on several different FLG flakes confirm the existence of flakes with varied thickness. The shape modification of 2D peak, which consists of four components (2D$_{1B}$, 2D$_{1A}$, 2D$_{2A}$, 2D$_{2B}$) provides information regarding the number of sheets in FLG (less and more than five) [20]. The broadening towards lower wavenumbers 2D peak (increase of relative intensity of 2D$_1$ component resulting in 2D splitting) indicates the presence of 3-4 layers containing flakes (Fig 3A). Fig. 3B, C and D shows also the $I_G$, $I_D$ and $I_D/I_G$ cartographies, respectively, which were performed within the singular flake presented on Fig. 3A. The Raman cartographies clearly show low (and homogenous) concentration of defects in the FLG plane, and the highest concentration of defects at the edges.

The FLG flakes exhibit large lateral size from few and up to dozens of microns. Such high surface allows to perform conductivity measurements on different and significant distance within a singular flakes. Such distance dependent electron transport measurements reflect the degree of continuity and homogeneity of the sample. Two examples of flakes, which are deposited between gold electrodes for the followed measurements, are presented in fig. 4 (top). The flakes are deposited randomly and the measurements are carried out on interdigitated FET-like device with two Au electrodes with gaps of 2.5, 5 and 10 μm. I(V) curves-measured for several flakes for each distance along with corresponding resistance values are given in fig 4. All I(V) curves are linear, revealing ohmic behavior. This indicates no (or negligible) scattering of charges due to the low defect content within the flakes, which is in agreement with the Raman analysis. The average resistance values obtained from few flakes are of 158, 238 and 365 Ω at the measured distance of 2.5, 5 and 10 μm, respectively.

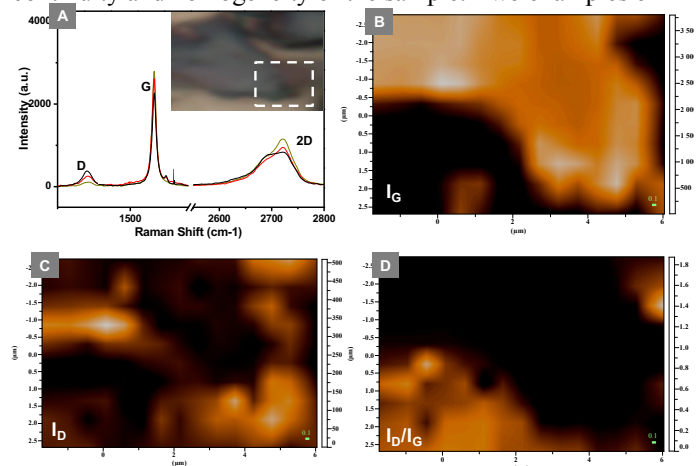

Figure 3. A) Representative Raman spectra (different flakes). B-D) $I_G$, $I_D$ and $I_D/I_G$ Raman cartographies obtained (calculated for $I_D/I_G$) form area of the flake presented at A.

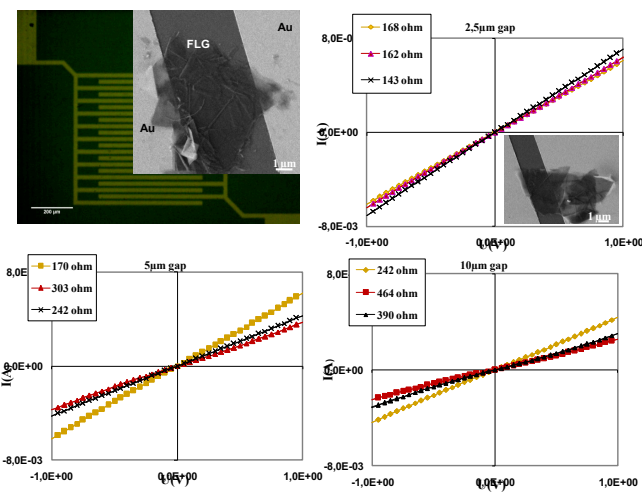

Figure 4. Representative images of measured FLG flake interdigitated FET-like devices and the I(V) curves obtained for few FLG flakes with varied electrodes distance (2.5, 5 and 10 μm), together with corresponding resistance values.

The linear increase of the resistance with a larger gap is due to the number of flakes, which are measured for each electrode distance (number of flakes is determined by optical microscopy). This average number decreases from ~ 10 for 2.5 μm, to ~7 for 5μm and to ~4 for 10 μm, which is related to higher probability of deposited flakes number within lower gap. Such deposited flakes behave as parallel resistors and the average resistance value for one flake for each electrodes distance can be found by multiplying the measured values (average) by the number of flakes, resulting in 1.58 kΩ, 1.67 kΩ and 1.46kΩ for 2.5, 5 and 10 μm respectively. The calculated minimum values for singular flakes are in the range of 500-700 Ω. The accordance of the average resistance values obtained for each electrodes distance is in agreement with the Raman cartographies revealing the homogeneity of the structure and the absence of significant defects in FLG flakes even at higher distance, such as 10 μm. The conductivity would be estimated. For the "square" flake with 7 sheets and resistance of 1.6 kΩ (thickness: $(7-1) \times 0.34 \times 10^{-9}$), a conductivity is of ~ $3 \times 10^5$ S/m, while for the flakes with the 500Ω resistance, it reaches almost $10^6$ S/m ($9.8 \times 10^5$ S/m). Such high conductivity is comparable with the conductivity parallel to the sheets (horizontal) ($\sigma_{\parallel}$), measured for graphite monocrystal or HOPG [21]. The electrical property of the FLG is also comparable with CVD-grown materials and superior to most of common exfoliation originated graphenes.

## CONCLUSION

FLG flakes with large lateral size and high electrical conductivity can be prepared by rapid exfoliation (extraction) of EG in toluene with the assistance of μ-waves irradiation, and their yield can be improved by performing a prior stirring of the suspension under low pressure. Significant absorption of μ-waves by EG/FLG allows the preparation of low defects, high range homogenous conductive material. Very low sheet resistance together with large lateral size of FLG flakes (up to tens of microns) are comparable to those of high quality graphenes synthesized by CVD method. Considerable sheets' size is of high interest to reach easy percolation in film formation [22].

## CONFLICT OF INTEREST

Higher Education Commission, Pakistan is acknowledged for student financial support (Azhar A. Pirzado). Other authors thank to Conectus program (Alsace Region).

## ACKNOWLEDGEMENTS

Thierry Romero is acknowledged for performing the SEM microscopy.